\documentclass[numberedappendix]{emulateapj} 
\newcommand{\bl}[1]{\mbox{\boldmath$ #1 $}}

\slugcomment{accepted for publication by The Astrophysical Journal}
\shorttitle{Properties of embedded  disks}
\shortauthors{Vorobyov} 

\begin{document}

\title{Embedded protostellar disks around (sub-)solar protostars. I. 
Disk structure and evolution}
\author{Eduard I. Vorobyov\altaffilmark{1,}\altaffilmark{2}}
\altaffiltext{1}{Institute for Computational Astrophysics, Saint Mary's University,
Halifax, B3H 3C3, Canada; vorobyov@ap.smu.ca.} 
\altaffiltext{2}{Research Institute of Physics, South Federal University, Stachki 194, Rostov-on-Don, 
344090, Russia.} 


\begin{abstract}
We perform a comparative numerical hydrodynamics study of embedded protostellar disks formed as 
a result of the gravitational collapse of cloud cores of distinct mass 
($M_{\rm cl}$=0.2--1.7~$M_\sun$) and ratio of rotational to gravitational
energy ($\beta$=0.0028--0.023). 
An increase in $M_{\rm cl}$ and/or $\beta$ leads to the formation of protostellar disks that are 
more susceptible to gravitational instability. Disk fragmentation occurs in most models but its 
effect is often limited to the very early stage,
with the fragments being either dispersed or driven onto the forming star during tens of orbital
periods. Only cloud cores with high enough $M_{\rm cl}$ or $\beta$ may eventually form 
wide-separation binary/multiple systems 
with low mass ratios and brown dwarf or sub-solar mass companions. It is feasible that 
such systems may eventually break up, giving birth to rogue brown dwarfs.
Protostellar disks of {\it equal} age formed from 
cloud cores of greater mass (but equal $\beta$) are generally denser, hotter, larger, 
and more massive. 
On the other hand, protostellar disks formed from cloud cores of higher $\beta$ (but equal $M_{\rm cl}$)
are generally thinner and colder but larger and more massive. In all models, the difference 
between the irradiation
temperature and midplane temperature $\triangle T$ is small, except for 
the innermost regions  of young disks, dense fragments, and disk's outer edge where $\triangle T$ 
is negative and may reach a factor of two or even more. Gravitationally unstable, embedded disks 
show radial pulsations, the amplitude of which increases along the line 
of increasing $M_{\rm cl}$ and $\beta$ but tends to diminish as the envelope clears. 
We find that single stars with a disk-to-star mass ratio of order unity can be formed only 
from high-$\beta$ cloud cores, but such massive disks are unstable and quickly fragment into 
binary/multiple systems. A substantial fraction of an embedded disk, 
especially its inner regions, spiral arms and dense clumps, may be optically thick, leading potentially
to observational underestimates of disk masses in the embedded phase of star formation.
\end{abstract}


\section{Introduction}
The early phases of the evolution of a protostellar disk, starting from its 
formation and ending with the clearing of a natal cloud core, determine the course along 
which a young stellar object (YSO) will evolve later in the T~Tauri phase. Vigourous gravitational 
instability that develops in protostellar disks during the embedded phase of star formation (EPSF)
serves to limit disk masses, effectively setting an upper limit 
on the mass accretion rates in the T~Tauri phase and flattening the accretion rate--stellar mass 
relation for solar-type stars \citep{VB09a}. Disk fragmentation triggered by gas
infall onto the disk from the collapsing cloud core
could lead to the formation of giant planets or brown dwarfs, thus shaping the subsequent 
physical properties of stellar and planetary systems \citep[e.g.][]{Boley09a,Boley09b,Machida10,VB10a}.
The large spread in mass accretion rates inferred for embedded YSOs in Perseus, 
Serpens, and Ophiuchus star-forming regions by \citet{Enoch09} can be accounted for
by gravitational instability and fragmentation of protostellar disks in the EPSF \citep{Vor09b}.
In addition, episodic accretion caused by disk fragmentation \citep{VB05,VB06,VB10b}
can explain the observed luminosity spread in H-R diagrams of several-Myr-old star-forming regions 
\citep{Baraffe09} and resolve the long standing ``luminosity problem'' \citep{Dunham10}, whereby 
protostars are underluminous compared to the accretion luminosity expected from analytical 
collapse calculations.

In spite of the pivotal role of the EPSF, our knowledge of protostellar disks in this phase is 
embarrassingly inadequate owing to difficulties with both observations and modeling.
Observing this phase has been difficult because disks are only visible at wavelengths where resolution
is poor. The added complication of the envelope
structure in the embedded systems makes an extraction of disk parameters from
interferometer data extraordinarily difficult with modern facilities. No wonder that most observations
of disk properties have been focused on T~Tauri disks \citep[e.g.][and many others]{Akeson02,Kitamura02,
Vicente05,Pietu07,Andrews07,Andrews09,Isella09} and considerably fewer attempted
studies of the embedded disks have been done so far  \citep[e.g.][]{Andrews05, Eisner05, Jorgensen09}.

Self-consistent numerical simulations of protostellar disks in the EPSF is no less difficult than observations
due to vastly changing spatial and temporal scales involved. The matter is that it is not sufficient
to just consider an {\it isolated} system with some presumed disk-to-star mass ratio. A self-consistent
treatment of the interaction of the star/disk system with the natal cloud core is of considerable 
importance for the disk physics and this inevitably requires solving for a much larger spatial volume
than in isolated systems. As a consequence, most information about
the early evolution of embedded protostellar disks was drawn from one-dimensional numerical 
studies of thin axisymmteric disks with some approximate prescription for disk gravity 
and, in some cases, with a solution for the vertical disk structure in the so-called 1+1D 
approximation \citep[e.g.][]{Lin90,Bell94,Nakamoto94,Hueso05,Visser09,Rice10}. 
Alternatively, analytic models of protostellar disks subject to intense mass loading 
have been constructed, which cover a wide parameter space and can yield important information 
than can later be compared against more focused numerical hydrodynamics
simulations \citep[e.g.][]{Kratter08,Clarke09,Krumholz10}.

Some interesting insight into the early disk
evolution, accretion history, and velocity structure of the infalling gas was 
also obtained in two-dimensional numerical simulations of axisymmetric disks 
\citep[e.g.][]{Yorke99,Boss01,Zhu09}. Unfortunately, such simulations
cannot describe self-consistently mass and angular momentum transport due to gravitational 
torques, which cannot develop in axisymmetric disks. To resolve this problem, some sort of 
effective viscosity is often invoked. An alternative two-dimensional approach involves solving for the
evolution of protostellar disks in the thin-disk approximation, which allows for an accurate
treatment of gravitational torques \citep[e.g.][]{Nelson00,Johnson03}. These studies revealed the importance
of disk cooling for the development of gravitational instability and fragmentation but suffered from
the lack of disk interaction with a natal cloud core in the EPSF.

An important step forward was done in a series of papers by \citet{VB05,VB06,VB09a,VB09b} and 
\citet{Vor09a,Vor09b,Vor10b} who
employed two-dimensional numerical simulations in the thin-disk approximation 
with an accurate treatment of disk self-gravity to model self-consistently the formation and evolution
of protostellar disks. These studies have demonstrated 
that mass loading onto the disk from the infalling envelope can bring about qualitatively new phenomena
such as the burst mode of accretion \citep{VB05,VB06,VB10b} and formation of giant planets on wide orbits
\citep{VB10a}. 
Such two-dimensional simulations allow us to consider the long-term evolution of a large number of
self-consistently formed protostellar disks at the expense of the detailed disk vertical structure.

First fully three-dimensional simulations, though suffering from insufficient numerical resolution,
have revealed important information about the vertical structure of the disks and their susceptibility
to gravitational instability and  fragmentation \citep[e.g.][]{Laughlin94, Burkert97,Boss98,Pickett00}.
With the advent of more powerful computers and techniques, new and more detailed 
three-dimensional numerical simulations of the formation and evolution of
protostellar disks have started to emerge \citep{Krumholz07,Boley09a,Boley09b,Kratter10,
Machida10}. Unfortunately, due to still enormous computational load involved, these studies often
suffer from restricted parameter space, limited temporal and spatial resolution, 
simplified treatment of the gas infall onto the disk, and/or the lack of detailed thermal physics. 
In addition, due to severe resolution limitations, 3D numerical simulations often 
resort to the use of sink particles to trace forming stellar and planetary objects, 
which inevitably involves introducing additional free parameters into the model.


This is the second paper in a series addressing the physical properties of {\it embedded}
protostellar disks. In the first paper \citep{VB10b}, we have mainly focused on the burst
mode of accretion that developes in fragmenting disks.
In this paper, we perform a comprehensive numerical hydrodynamics study of the formation 
and long-term evolution of protostellar disks formed from cloud cores
of distinct mass and rotational energy. We make use of the thin-disk approximation 
complemented by a detailed calculation of the disk thermal balance. 
Several models were scrutinized to derive the detailed 
disk structure and time evolution. Our numerical studies can serve as a framework for the future 
high-resolution observations of embedded disks using such ground-based or space-based 
facilities as ALMA or Millimetron \citep{Wild09}. The paper is organized
as follows. In \S~\ref{model}, we review our numerical model and initial setup. The structure
and evolution of protostellar disks fromed from cloud cores of distinct 
mass and rotational energy are presented in \S~\ref{incrmass} and \S~\ref{increnergy}.
The main conclusions are summarized in \S~\ref{conclude}.

\section{Model description}
\label{model}
The main concepts of our numerical approach are explained in detail in \citet{VB10b}
and, for the reader's convenience, are briefly reviewed below.
We start our numerical integration in the pre-stellar phase, which is 
characterized by a collapsing {\it starless} cloud core, 
continue into the embedded phase of star formation, during which
a star, disk, and envelope are formed, and terminate our simulations in the T Tauri phase,
when most of the envelope has accreted onto the forming star/disk system.
In the EPSF, the disk occupies the innermost region of our numerical grid, while the 
larger outer part of the grid is taken up by the infalling envelope, 
the latter being the remnant of the parent cloud core. 
This ensures that the protostellar disk is not isolated in the EPSF but is exposed to intense
mass loading form the envelope. In addition, the mass accretion rate onto 
the disk $\dot{M}_{\rm env}$ is not a free parameter of the model 
but is self-consistently determined by the gas dynamics in the envelope.

We introduce a ``sink cell'' at $r_{\rm sc}=6$~AU and impose a free inflow inner boundary condition
and free outflow outer boundary condition so that that the matter is allowed to flow out of 
the computational
domain but is prevented from flowing in. We monitor the gas surface density in the sink cell and 
when its value exceeds a critical value for the transition from 
isothermal to adiabatic evolution, we introduce a central point-mass star.
In the subsequent evolution, 90\% of the gas that crosses the inner boundary 
is assumed to land onto the central star plus the inner axisymmetric disk at $r<6$~AU. 
This inner disk is dynamically inactive; it contributes only to the total gravitational 
potential and is used to secure a smooth 
behaviour of the gravity force down to the stellar surface.
The other 10\% of the accreted gas is assumed to be carried away with protostellar jets. 
The latter are triggered only after the formation of the central star.

\subsection{Basic equations}
We make use of the thin-disk approximation to compute the gravitational collapse of rotating, 
gravitationally unstable cloud cores. This approximation is an excellent means to calculate
the evolution for many orbital periods and many model parameters. 
The basic equations of mass, momentum, and energy transport  are
\begin{equation}
\label{cont}
\hskip -5 cm \frac{{\partial \Sigma }}{{\partial t}} =  - \nabla_p  \cdot 
\left( \Sigma \bl{v}_p \right),  
\end{equation}
\begin{eqnarray}
\label{mom}
\frac{\partial}{\partial t} \left( \Sigma \bl{v}_p \right) &+& \left[ \nabla \cdot \left( \Sigma \bl{v_p}
\otimes \bl{v}_p \right) \right]_p =   - \nabla_p {\cal P}  + \Sigma \, \bl{g}_p + \\ \nonumber
& + & (\nabla \cdot \mathbf{\Pi})_p, 
\label{energ}
\end{eqnarray}
\begin{equation}
\frac{\partial e}{\partial t} +\nabla_p \cdot \left( e \bl{v}_p \right) = -{\cal P} 
(\nabla_p \cdot \bl{v}_{p}) -\Lambda +\Gamma + 
\left(\nabla \bl{v}\right)_{pp^\prime}:\Pi_{pp^\prime}, 
\end{equation}
where subscripts $p$ and $p^\prime$ refers to the planar components $(r,\phi)$ 
in polar coordinates, $\Sigma$ is the mass surface density, $e$ is the internal energy per 
surface area, 
${\cal P}=\int^{Z}_{-Z} P dz$ is the vertically integrated
form of the gas pressure $P$, $Z$ is the radially and azimuthally varying vertical scale height
determined in each computational cell using an assumption of local hydrostatic equilibrium,
$\bl{v}_{p}=v_r \hat{\bl r}+ v_\phi \hat{\bl \phi}$ is the velocity in the
disk plane, $\bl{g}_{p}=g_r \hat{\bl r} +g_\phi \hat{\bl \phi}$ is the gravitational acceleration 
in the disk plane, and $\nabla_p=\hat{\bl r} \partial / \partial r + \hat{\bl \phi} r^{-1} 
\partial / \partial \phi $ is the gradient along the planar coordinates of the disk. 

Viscosity enters the basic equations via the viscous stress tensor $\mathbf{\Pi}$ expressed as
\begin{equation}
\mathbf{\Pi}=2 \Sigma\, \nu \left( \nabla \bl{v} - {1 \over 3} (\nabla \cdot \bl{v}) \mathbf{e} \right),
\label{stressT}
\end{equation}
where $\mathbf{e}$ is the unit tensor. We note that we take no simplifying
assumptions about the form of $\mathbf{\Pi}$ apart from those imposed by the adopted thin-disk 
approximation. 
We parameterize the magnitude of kinematic viscosity $\nu$ using a modified form 
of the $\alpha$-prescription 
\begin{equation}
\nu=\alpha \, c_{\rm s} \, Z \, {\cal F}_{\alpha}(r), 
\end{equation}
where $c_{\rm s}^2=\gamma {\cal P}/\Sigma$ is the square of effective sound speed
calculated at each time step from the model's known ${\cal P}$ and $\Sigma$. The 
function ${\cal F}_{\alpha}(r)=
2 \pi^{-1} \tan^{-1}\left[(r_{\rm d}/r)^{10}\right]$ is a modification to the usual 
$\alpha$-prescription that guarantees that the turbulent viscosity operates 
only in the disk and quickly reduces to zero beyond the disk radius $r_{\rm d}$.
In this paper, we use a spatially and temporally 
uniform $\alpha$, with its value set to $5\times 10^{-3}$ based on our recent work \citep{VB09b}. 
Our adopted value of $\alpha$ are in agreement with the mean value inferred by \citet{Andrews09} for
a large sample of protostellar disks in the Ophiuchus star-forming region.

Equation~(\ref{energ}) for the internal energy (per surface area) transport includes 
compressional heating ${\cal P}\left( \nabla_p \cdot \bl{v}_p \right)$, radiative cooling 
$\Lambda$, heating due to stellar/background irradiation $\Gamma$, 
and viscous heating $(\nabla \bl{v})_{pp^\prime}:\Pi_{pp^\prime}$. 
The cooling function is determined using the diffusion
approximation of the vertical radiation transport in a one-zone model of the vertical disk 
structure \citep{Johnson03}
\begin{equation}
\Lambda={\cal F}_{\rm c}\sigma\, T^4 \frac{\tau}{1+\tau^2},
\end{equation}
where $\sigma$ is the Stefan-Boltzmann constant, $T$ is the midplane temperature of gas, 
and ${\cal F}_{\rm c}=2+20\tan^{-1}(\tau)/(3\pi)$ is a function that 
secures a correct transition between the cooling function (from both surfaces of the disk) 
in the optically thick regime $\Lambda_{\rm thick}=16\, \sigma \, T^4/3\tau$ 
and the optically thin one $\Lambda_{\rm thin}=2\,\sigma \,T^4\,\tau$.  We use 
frequency-integrated opacities of \citet{Bell94}

The heating function is expressed as
\begin{equation}
\Gamma={\cal F}_{\rm c}\sigma\, T_{\rm irr}^4 \frac{\tau}{1+\tau^2},
\end{equation}
where $T_{\rm irr}$ is the irradiation temperature at the disk surface 
determined by the stellar and background black-body irradiation as
\begin{equation}
T_{\rm irr}^4=T_{\rm bg}^4+\frac{F_{\rm irr}(r)}{\sigma},
\label{fluxCS}
\end{equation}
where $T_{\rm bg}$ is the uniform background temperature (in our model set to the 
initial temperature of the natal cloud core)
and $F_{\rm irr}(r)$ is the radiation flux (energy per unit time per unit surface area) 
absorbed by the disk surface at radial distance 
$r$ from the central star. The latter quantity is calculated as 
\begin{equation}
F_{\rm irr}(r)= A_{\rm irr}\frac{L_\ast}{4\pi r^2} \cos{\gamma_{\rm irr}},
\end{equation}
where $\gamma_{\rm irr}$ is the incidence angle of 
radiation arriving at the disk surface at radial distance $r$ and  $A_{\rm irr}=M_{\rm cl}/
(M_{\rm env}+M_{\rm cl})$ is a time-dependent
factor that accounts for the attenuation of stellar radiation by the envelope with mass $M_{\rm env}$
in three-dimensional disks and takes values from $\ga0.5$ at the disk formation to $\approx 1.0$ at the end of the EPSF and beyond\footnote{Although most of the 
envelope material should be landing onto the disk outer edge, a smaller fraction 
may still be falling onto the disk inner regions \citep{Visser09} and 
this material may intercept some of the stellar radiation in the EPSF. 
The effect of this attenuation factor on the disk evolution is however not significant,
since its value quickly approaches unity. For instance in model~3, $A_{\rm irr}\approx0.75$ at 
$t=0.1$~Myr after the disk formation and $A_{\rm irr}\approx0.95$ at $t=0.3$~Myr.
}. 
The stellar luminosity $L_\ast$ is the sum of the accretion luminosity $L_{\rm \ast,accr}=G M_\ast \dot{M}/2
r_\ast$ arising from the gravitational energy of accreted gas and
the photospheric luminosity $L_{\rm \ast,ph}$ due to gravitational compression and deuterium burning
in the star interior. The stellar mass $M_\ast$ and accretion rate onto the star $\dot{M}$
are determined self-consistently during numerical simulations via the amount of gas passing through
the sink cell. The stellar radius $r_\ast$ is calculated using an approximation formula of \citet{Palla91},
modified to take into account the formation of the first molecular core \citep{Masunaga00}.
The photospheric luminosity $L_{\rm \ast,ph}$ is taken from the pre-main 
sequence tracks for the low-mass stars and brown dwarfs calculated by \citet{DAntona97}. 
More details are given in \citet{VB10b}.

Viscous heating operates in the disk interior and is calculated using the standard expression 
$(\nabla \bl{v})_{pp^\prime} : \mathbf{\Pi}_{pp^\prime}$. 
Heating due to shock waves is taken into account via compressional heating 
${\cal P}\left(\nabla_p \cdot \bl{v}_p \right)$ and 
artificial viscosity, the latter implemented in the code using the standard prescription 
of \citet{RM57}. 
The vertically integrated gas pressure ${\cal P}$ and internal energy per surface area $e$ are 
related via the ideal 
equation of state ${\cal P}=(\gamma-1)\, e$, with the ratio of specific heats $\gamma=7/5$.

Equations~(\ref{cont})--(\ref{energ}) are solved using the method of finite differences with a time
explicit solution procedure in polar coordinates $(r, \phi)$ on a numerical grid with
$512 \times 512$ grid zones. The advection is treated using the van Leer interpolation scheme.
The update of the internal energy per surface area 
$e$ due to cooling $\Lambda$ and heating $\Gamma$
is done implicitly using the Newton-Raphson method of root finding, complemented by the bisection method
where the Newton-Raphson iterations  fail to converge. 
The viscous heating and force terms in Equations~(\ref{mom}) and (\ref{energ}) are 
implemented in the code  using an explicit finite-difference
scheme, which is found to be adequate for $\alpha\la 0.01$.
The radial points are logarithmically spaced.
The innermost grid point is located at the position of the sink cell $r_{\rm sc}=6$~AU, and the 
size of the first adjacent cell varies in the 0.07--0.1~AU range depending on the cloud core 
size.  This corresponds to the radial resolution of $\triangle r$=1.1--1.6~AU at 100~AU.

\subsection{Initial conditions}
Initially isothermal ($T_{\rm init}\equiv T_{\rm bg}=10$~K) cloud cores have surface densities 
$\Sigma$ and angular velocities $\Omega$ typical for a collapsing, axisymmetric, magnetically
supercritical core \citep{Basu97}
\begin{equation}
\Sigma={r_0 \Sigma_0 \over \sqrt{r^2+r_0^2}}\:,
\label{dens}
\end{equation}
\begin{equation}
\Omega=2\Omega_0 \left( {r_0\over r}\right)^2 \left[\sqrt{1+\left({r\over r_0}\right)^2
} -1\right],
\label{omega}
\end{equation}
where $\Omega_0$ is the central angular velocity and 
$r_0$ is the radius of central near-constant-density plateau defined 
as $r_0 = \sqrt{A} c_{\rm s}^2 /(\pi G\Sigma_0)$, with the initial positive density enhancement $A$
set to 1.2 throughout the paper. Model cores are characterized by a distinct ratio 
$r_{\rm out}/r_0=6$ in order to generate gravitationally unstable truncated cores of similar 
form, where $r_{\rm out}$ is the cloud core's outer radius. 

\begin{table}
\begin{center}
\caption{Model cloud core parameters}
\label{table1}
\begin{tabular}{cccccc}
\hline\hline
Model & $M_{\rm cl}$ & $\beta$ & $\Omega_0$ & $r_0$ & $\Sigma_0$   \\
\hline
 1 & 0.2  & $5.6\times 10^{-3}$ & 6.2 & 445  & 0.28  \\
 2 & 0.85 & $5.6\times 10^{-3}$ & 1.5 & 1885 & 0.066 \\
 3 & 1.7  & $5.6\times 10^{-3}$ & 0.7 & 3770 & 0.033 \\
 4 & 0.85 & $2.8\times 10^{-3}$ & 1.0 & 1885 & 0.066 \\
 5 & 0.85 & $2.3\times 10^{-2}$ & 2.9 & 1885 & 0.066 \\
 \hline
\end{tabular} 
\tablecomments{All masses are in $M_\sun$, distances in AU, surface densities in g~cm$^{-2}$, and 
angular velocities in km~s$^{-1}$~pc$^{-1}$.}
\end{center}
\end{table} 

For the in-depth analysis of circumstellar disk structure and evolution, we have chosen 
five typical model cloud cores with varying initial cloud core masses $M_{\rm cl}$ 
and ratios $\beta=E_{\rm rot}/|E_{\rm grav}|$ of the rotational to gravitational  energy 
defined as
\begin{equation}
E_{\rm rot}= 2 \pi \int \limits_{r_{\rm sc}}^{r_{\rm
out}} r a_{\rm c} \Sigma \, r \, dr,
\label{rotEn}
\end{equation}
\begin{equation}
E_{\rm grav}= - 2\pi \int \limits_{r_{\rm sc}}^{\rm r_{\rm out}} r
g_r \Sigma \, r \, dr,
\label{gravEn}
\end{equation}
respectively, 
where $a_{\rm c} = \Omega^2 r$ is the  
centrifugal acceleration. The parameters of these models are listed in Table~\ref{table1}. We note that
cloud cores with the adopted values of $M_{\rm cl}$ are expected to form solar- and sub-solar type stars
and the values of $\beta$ lie within the limits inferred by \citet{Caselli} for dense molecular
cloud cores, $\beta=(10^{-4} - 0.07)$.

In the following text, we consider separately models with distinct $M_{\rm cl}$ but similar $\beta$
(models 1, 2, and 3) and distinct $\beta$ but similar $M_{\rm cl}$ (models~2, 4, and 5). 
In particular, the parameter $\beta$ is varied by changing the value of $\Omega_0$ only.
These two sets of models have been chosen to single out the effects 
of increasing cloud core masses, from one hand, and increasing rotational energies of cloud cores, 
from the other hand. We note that the original cloud core mass and rotation rate
are expected to have a major impact on the evolution of a protostellar disk because 
they determine the centrifugal radius and, consequently, the disk mass and stability properties.

\section{Disk structure and evolution along the line of increasing cloud core masses}
\label{incrmass}
We start by considering the properties of protostellar disks formed from cloud cores of 
increasingly higher mass $M_{\rm cl}$. In particular, model~1 is characterized 
by $M_{\rm cl}=0.2~M_\sun$, while models~2 and 3 have $M_{\rm cl}=0.85~M_\sun$ and 
$M_{\rm cl}=1.7~M_\sun$, respectively. The ratio $\beta$ is identical for these models 
and is set to $5.6\times 10^{-3}$.
\label{diskstructure}

\begin{figure*}
 \centering
  \includegraphics[width=16cm]{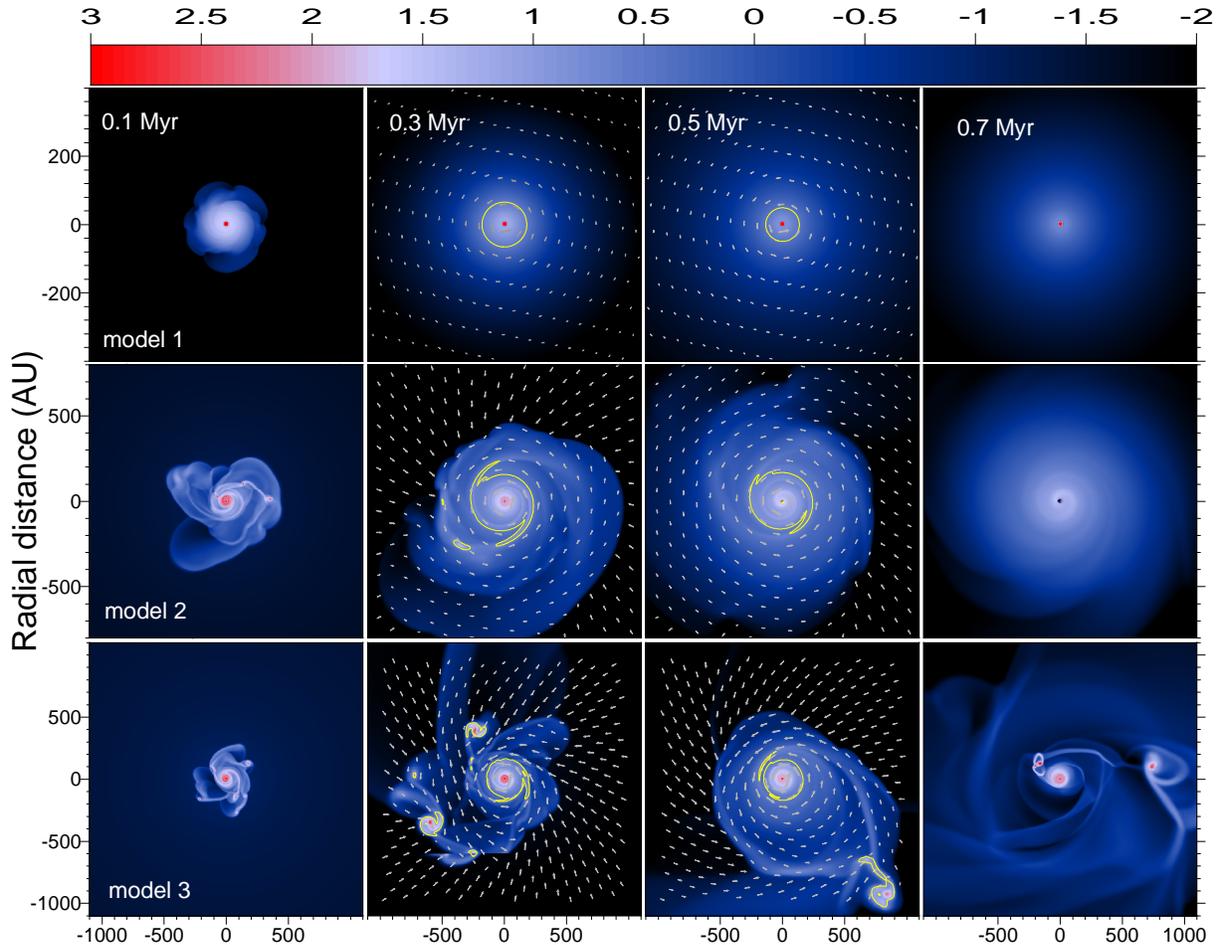}
      \caption{Gas surface density (in g~cm$^{-2}$, log units) in three models with distinct
      cloud core mass $M_{\rm cl}$ and at distinct times $t$ after the disk formation. 
      In particular, the top row  shows disk images in the $M_{\rm cl}=0.2~M_\sun$
      model, while the middle and bottom rows present those of the $M_{\rm cl}=0.85~M_\sun$
      and $M_{\rm cl}=1.7~M_\sun$ models, respectively. Columns from left to right
      correspond to $t=0.1$~Myr, $t=0.3$~Myr, $t=0.5$~Myr, and $t=0.7$~Myr, respectively. 
      Superimposed on the disk
      images are the gas velocity fields (arrows) and  yellow contour lines delineating disk regions
      with the frequency-integrated optical depth $\tau\ge0.1$.}
         \label{fig1}
\end{figure*}

\subsection{Face-on disk images}
Figure~\ref{fig1} presents gas surface densities (in g~cm$^{-2}$, log units) in model~1 (top row),
model~2 (middle row) and model~3 (bottom row) at four distinct times after the disk formation (from
left to right): $t$=0.1~Myr, $t$=0.3~Myr, $t$=0.5~Myr, and $t=0.7$~Myr. 
We note that the spatial scale is different for model~1 ($400 \times 400$~AU), model~2 ($800 \times
800$~AU), and model~3 ($1100 \times 1100$~AU). Superimposed on
the gas surface density images are the gas velocity fields (arrows) and yellow contour lines. 
The latter delineate disk regions that are characterized by the frequency-averaged 
optical depth $\tau\ge0.1$. Optical depth effects are expected to become important in these regions.

It is evident that the disk appearance in Figure~\ref{fig1} changes significantly  
along both the line of increasing cloud core mass (from top to bottom) and 
the line of growing disk age (from left to right).
The disk in the $M_{\rm cl}=0.2~M_\sun$ model~1 (top row) 
demonstrates a weak, diffuse spiral 
structure at $t=0.1$~Myr and becomes nearly perfectly axisymmetric in the subsequent evolution. 
On the other hand, the disk in the $M_{\rm
cl} =0.85~M_\sun$ model~2 (middle row) exhibits a rich, well-defined spiral structure 
at $t=0.1$~Myr, 
which weakens and diffuses with time but still can be traced in the 0.5-Myr-old disk. 
Fragmentation of the spiral arms (small red clumps in the middle-left panel) 
occurs at $t$=0.1~Myr after the disk formation. However, no survived fragments are seen at later 
times due to an efficient migration mechanism\footnote{See animation of this quick migration 
at www.astro.uwo.ca/$\sim$vorobyov 
(animations: burst mode of accretion).} that operates in the EPSF and drives the 
fragments into the inner disk regions and through the sink cell, causing accretion and luminosity 
bursts \citep{VB05,VB06,VB10b}. An apparent lopsidedness is evident in the 0.3-Myr-old
and 0.5-Myr-old disks but seems to diminish at later times. The gas 
velocity field indicates considerable non-circular motions within the disk, suggesting the 
existence of local deviations from a Keplerian rotation. In addition, notable non-axisymmetric
variations in the disk shape suggest that an axisymmetric description
of a circumstellar disk in the early evolution stage is unjustified, at least on time scales
of the order of 0.5~Myr after the disk formation.

\begin{figure}
  \resizebox{\hsize}{!}{\includegraphics{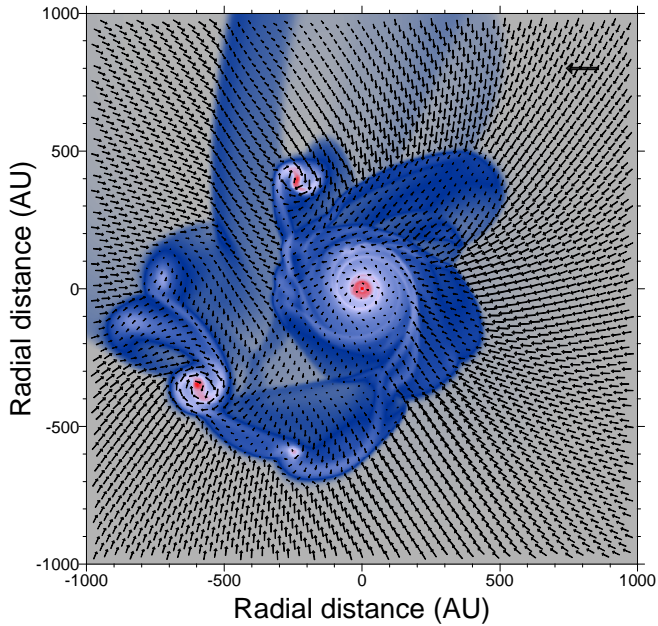}}
      \caption{Residual velocity field (total velocity field minus Keplerian rotation) 
      superimposed on the gas surface density in model~3 at t=0.3~Myr
      after the disk formation. The horizontal arrow in the right-upper corner has a dimension of 5~km~s$^{-1}$.
      In the simulation, the disk rotates counterclockwise and two fragments with disk rotating clockwise
      are clearly visible. }
         \label{fig2}
\end{figure}

The $M_{\rm cl}=1.7$ model~3 (bottom row in Figure~\ref{fig1}) is most extreme and 
demonstrates vigourous gravitational instability and fragmentation. Two well-defined fragments
possessing counterrotating minidisks and a few smaller ones are seen in the 0.3-Myr-old disk
(note that the disk rotates counterclockwise). 
The counterrotation can clearly be seen in Figure~\ref{fig2}, which shows the residual 
velocity field
superimposed on the gas surface density in model~3 at $t=0.3$~Myr. 
The residuals are obtained by subtracting the Keplerian rotation from the total gas velocity, 
thus accentuating the gas motion in a local frame of reference moving with the disk.
Considerable non-Keplerian motions are also evident near/at the spiral arms. 

Such counterotating structures may owe their existence to differential rotation 
of fragmenting spiral arms. When a fragment forms near corotation\footnote{Fragmentation 
of spiral arms occurs preferentially near corotation where distortion due to differential 
rotation is minimal.}, it captures gravitationally some of 
the neighboring material, which receives a counterrotating twist around the forming fragment 
due to the fact that the inner parts of the arm rotate around the central star faster than 
the outer ones. It is important to note that this mechanism is not universal and,
as found in other runs \citep{VB10b}, some fragments may also possess corotating minidisks. 
It is feasible that the direction of rotation is determined by local conditions at the position 
of the forming
fragment (e.g. spiral arm structure and density, residual velocity field, etc.), 
in which case it is difficult to predict the outcome. 
We plan to study this phenomenon in greater detail in a future paper.

In the 0.5-Myr-old disk, most of the fragments have either been destroyed or absorbed 
by the central star. However, one fragment has survived and a binary system 
with the smaller companion possessing a sizeable counterotating disk has emerged. 
The separation between the companions is about 1000~AU and the masses of the primary/secondary are 
approximately $M_\ast=0.95~M_\sun$ and $M_{\rm sec}=0.15~M_\sun$, respectively, with the latter 
value being an upper limit on both the companion and its minidisk. 
The fragmentation process continues even in the 0.7-Myr-old system
and another fragment with mass $\le0.08~M_\sun$ emerges in the disk of the primary at $r\approx200$~AU.
In the meantime, the outer fragment has migrated inward to approximately 720~AU.
The subsequent evolution of the system is uncertain since we terminate the simulation 
due to enormous computational load. We hope that model~3 can ultimately evolve into a binary or triple
system with low-mass ratio and (at least) one of the companion being in the brown dwarf mass regime.

Three interesting features in Figure~\ref{fig1} are worth of specific emphasis.
First, young protostellar disks ($t\la0.1$~Myr) are conspicuously non-axisymmetric,
even in the $M_{\rm cl}=0.2~M_\sun$ model, but evolve with time toward a more regular, axisymmetric
state. However, the rate of this evolution is much faster in models with lower cloud core masses. This
is primarily caused by the fact that mass loading from the infalling envelope is the main driving 
force that sustains gravitational instability in the disk and the duration of this infall 
(i.e., the lifetime of the EPSF) is near-linearly proportional to the cloud core mass \citep{Vor10b}.
Second, a fairly large portion of the disk, irrespective of the model, 
is characterized by $\tau>0.1$, which implies that the optical depth effect cannot be neglected
in the early evolution of protostellar disks. We will return to this question in more
detail in \S~\ref{radstructure}.
Finally, sufficiently massive cloud cores may form binary or multiple systems with 
brown dwarf and/or sub-solar companions. In the future higher-resolution simulations, 
we hope to see planetary-mass objects forming in the disk.

\begin{figure*}
 \centering
  \includegraphics[width=15cm]{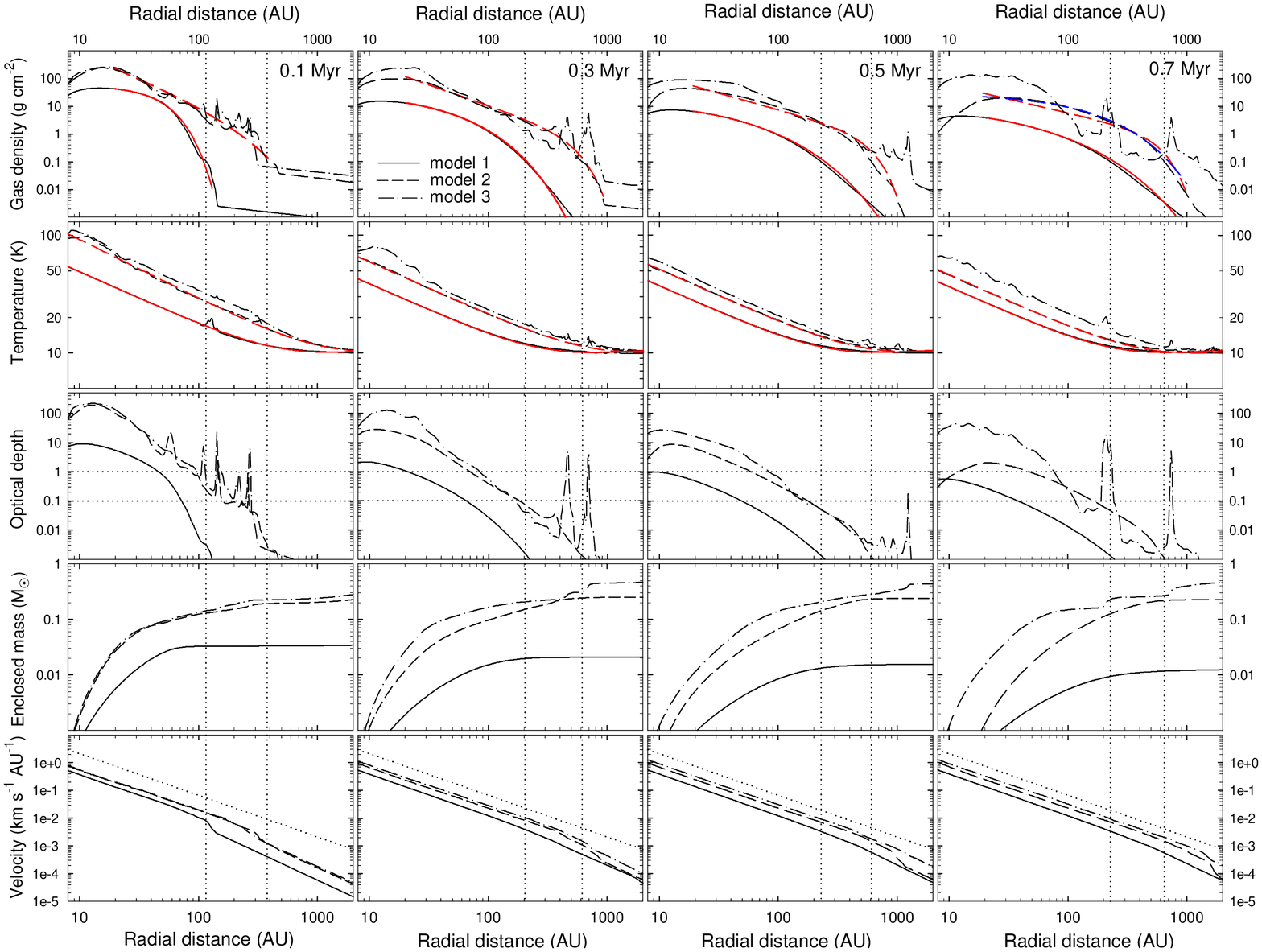}
      \caption{Azimuthally-averaged radial profiles of the gas surface density
      $\overline \Sigma$ (first row), gas midplane temperature $\overline T$ (second row), 
      frequency-integrated optical depth $\overline \tau$ (third row), and angular velocity 
      $\overline \Omega$ (fifth row). The forth row presents the enclosed mass $M(r)$ as a function
      of radius. The vertical columns from left to right show the data at different times after the
      disk formation: $t$=0.1~Myr, $t$=0.3~Myr, $t=0.5$~Myr, and $t=0.7$~Myr.
      The solid, dashed, and dash-dotted lines present the data for the $M_{\rm cl}=0.2~M_\sun$ model~1,
      $M_{\rm cl}=0.85~M_\sun$ model~2, and $M_{\rm cl}=1.7~M_\sun$ model~3, respectively. The red solid
      and dashed lines are the best fits to the model gas surface density and midplane temperature
      distributions in model~1 and model~2, respectively. The vertical dotted lines mark the position
      of the disk outer edge in model~1 (left lines) and model~2 (right lines). The horizontal dotted
      lines are the characteristic optical depths of $\tau=1.0$ and $\tau=0.1$. The sloped dotted lines
      in the bottom row show the angular velocity of a pressure-free gas in the gravitational field
      of a 5.0~$M_\sun$ star.}
         \label{fig3}
\end{figure*}

\subsection{Disk radial structure}
\label{radstructure}
Due to limited spatial resolution of modern observational facilities at long wavelengths, 
vital information about the structure of protostellar disks is often retrieved from 
modeling the spectral 
energy distribution based on analytical disk models that assume gas surface densities 
and temperatures to be  power laws in radius \citep[e.g.][]{Looney03,Andrews05,Andrews09}. 
Our numerical simulations  can provide a useful starting point for such an analysis
by constructing the typical azimuthally-averaged radial profiles of various 
physical characteristics of protostellar disks.

Figure~\ref{fig3} presents such radial profiles for the gas surface density
$\overline \Sigma$ (first row), midplane gas temperature $\overline T$ (second row), 
optical depth $\overline \tau$ (third row), enclosed mass $M(r)$ (forth row), 
and angular velocity $\overline \Omega$ (fifth row). In particular, black solid lines correspond to
model~1, while black dashed and dash-dotted lines present the data for models~2 and 3, respectively.
The vertical columns from left to right correspond to four distinct times after the disk formation:
$t=0.1$~Myr, $t$=0.3~Myr, $t$=0.5~Myr, and $t=0.7$~Myr.
The vertical dotted lines show the disk radii $r_{\rm d}$ in model~1 (left line) and
model~2 (right line) as derived from a characteristic gas surface density of 
$\Sigma_{\rm d2e}=0.1$~g~cm$^{-2}$ for the disk to envelope transition and radial gas velocity 
\citep[see][for details]{Vor10b}.
The horizontal dotted lines in the third row mark characteristic optical depths of 
$\overline \tau$=1 (top) and $\overline\tau$=0.1 (bottom).  The sloped dotted lines in 
the bottom row show the angular velocity of a
pressure-free gas in the gravitational field of a 5.0~$M_\sun$ star. The red solid and dashed
lines and also the blue dashed line are the best fits to the model gas density and temperature distributions.

The top row in Figure~\ref{fig3} demonstrates that protostellar disks of {\it equal} age formed
from cloud cores of greater mass are generally characterized by higher gas surface densities 
$\overline \Sigma$, larger radii $r_{\rm d}$, and, consequently, greater masses $M_{\rm d}$.
For instance, the 0.1-Myr-old disk in model~1 has  $\overline \Sigma(r=20~\mathrm{AU})=42$~g~cm$^{-2}$,
$r_{\rm d}$=117~AU, and $M_{\rm d}=0.03~M_\sun$, while the 
corresponding values in model~2 are $\overline \Sigma(r=20~\mathrm{AU})=208$~g~cm$^{-2}$, 
$r_{\rm d}$=378~AU, and $M_{\rm d}=0.2~M_\sun$. 
This tendency is a mere consequence of an increased mass reservoir in the envelope, increased mass
infall rate onto the disk, and insufficiently fast radial mass transport in low-mass disks 
dominated by viscous torques with typical $\alpha$ of the order of
0.005--0.01. As a result, low-mass disks cannot efficiently transport matter from
the disk's outer edge (where most of the envelope gas lands) onto the forming star and 
the disk starts to grow both in mass and radius. A similar tendency of T~Tauri disks being in general
denser and hotter than those of substellar objects was also found by \citet{Wiebe08}.

However, the trend of more massive cloud cores to form denser disks becomes much 
less pronounced for cores with $M_{\rm cl}\ga 0.9~M_\sun$.
Indeed, the azimuthally-averaged gas surface density in model~3 is only marginally 
higher than in model~2,  
indicating the onset of a self-regulatory state in which the disk grows in size and mass 
(due to continuing mass loading from the infalling envelope)  rather than in gas surface density. 
This effect is caused by vigourous gravitational instability and 
fragmentation that develop in the EPSF.
Indeed, as Figure~\ref{fig1} demonstrates, the $M_{\rm cl}=0.2~M_\sun$ model~1 reveals a rather 
weak spiral structure, while the protostellar disk in the $M_{\rm cl}=1.7~M_\sun$ model~3 is 
conspicuously  non-axisymmetric, showing a well-defined spiral structure and multiple 
fragments forming in the arms. These fragments migrate radially inward and onto the star
during a few tens of orbital periods (often on even shorter timescales), 
thus effectively restricting the growth of the disk surface density and 
bringing the disk back to the border of stability after every fragmentation episode.


The behaviour of both the midplane gas temperature (second row in Figure~\ref{fig3}) 
and optical depth (third row) in disks of equal age is similar to that of the gas surface 
density (top row)---both $\overline T$ and $\overline \tau$ increase along the sequence 
of increasing cloud core masses. As in the case with $\overline\Sigma$, 
this trend appears to slow down for models with $M_{\rm cl}\ga 0.9~M_\sun$.

From the third row in Figure~\ref{fig3} it is seen that the inner disk regions and most fragments 
are optically thick with $\overline{\tau}>1.0$, while the outer disk regions are either marginally 
optically thin ($0.1<\overline{\tau}<1.0$)  or completely optically thin ($\overline{\tau}<0.1$). 
The fact that a substantial
fraction of protostellar disks is optically thick may have important consequences for the disk mass
estimates using conventional observational techniques. Indeed, the forth row in Figure~\ref{fig3} 
presents the enclosed mass $M(r)$ (both that of the disk and the envelope) as a function of 
radius. It is evident
that a considerable fraction of the total mass may be concentrated in regions with $\overline{\tau}>0.1$,
which are expected to be affected by a non-negligible optical depth. This is particularly true 
for the early disk evolution.
For instance, the 0.3-Myr-old disk in model~2 has radius $r_{\rm d}\approx 615$~AU and mass 
$M_{\rm d}=0.24~M_\sun$.
On the other hand, if we count only those regions that are characterized by $\overline{\tau}>0.1$ 
and $\overline{\tau}>1$, the corresponding disk masses are 0.075~$M_\sun$ and 0.14~$M_\sun$, 
respectively, indicating that as much as 70\% of the total 
mass content may be hidden from our sight. 
As Figure~\ref{fig3} demonstrates, both $\overline \Sigma$ and $\overline \tau$ express a tendency 
to decrease with time, implying 
that the observationally inferred disk mass are expected to be more accurate 
for disks of older age. 

The bottom row in Figure~\ref{fig3} reveals that the {\it azimuthally-averaged} angular velocity 
has a near-Keplerian profile,
notwithstanding conspicuous local deviations from a circular gas motion evident in Figure~\ref{fig1}.
In addition, the outer disk regions exhibit a slightly over-Keplerian rotation, most
likely due to an additive effect from the gravitational field of the inner disk regions. 

It is useful to construct some functions that approximate and summarize our model radial profiles
of $\overline \Sigma$ and $\overline T$ for disks of distinct age. 
To fit our model data, we use the following two functions
\begin{eqnarray}
\label{func1}
{\overline\Sigma} &= &\Sigma_0 \left( {\mathrm{1 AU} \over r} \right)^{n_1} 
\exp\left[{-\left( r\over a \right)^{m}}\right],\\
\label{func2}
{\overline T} &=& T_0 \left( {\mathrm{1 AU} \over r } \right)^{n_2} +
T_{\rm bg} \left( {r\over r+b} \right),
\end{eqnarray}
where $r$ is the radial distance in AU, $T_0$ and $\Sigma_0$ are the gas temperature (in Kelvin) 
and gas surface density (in g~cm$^{-2}$) at 1~AU, respectively\footnote{The parameters $a$ and $b$ 
are usually large numbers.}. The form of Equation~(\ref{func1}) is chosen to approximate $\overline
\Sigma$ in the disk inner and intermediate regions by a power-law function, while
the disk outer regions, which are usually characterized by gas surface density falling off 
with radius faster than a power-law, are approximated by an exponential function. However, if 
$n_1$ is set to zero, Equation~(\ref{func1}) approximates the whole disk by a sole 
exponential function. In the case of $n_1\ne0$, the parameters $m$ and $a$ express 
the steepness of $\overline \Sigma$ near 
the disk outer edge and the approximate radial position of the disk outer edge, respectively. 
In the case of $n_1=0$, the aforementioned parameters express the steepness and the characteristic radius
of the radial gas surface density distribution in the whole disk.
The form of Equation~(\ref{func2}) is motivated by the fact that the radial gas temperature 
distribution in the midplane of a flared disk is usually well approximated by a power-law function,
while the right-hand-side term in Equation~(\ref{func2}) is invoked to make a smooth transition
between the disk and the envelope, the latter being characterized by a background temperature 
$T_{\rm bg}$.
The parameter $b$ represents the radial position beyond which the
gas thermal balance is controlled by the external background radiation with temperature 
$T_{\rm bg}$ rather than by stellar irradiation or viscous heating. 

Red lines in the upper two rows of Figure~\ref{fig3} present our best fits
to the model radial distributions of $\overline \Sigma$ and $\overline T$. More specifically,
the red solid and dashed lines are the best fits to model~1 and 2, respectively. 
We do not fit model~3, since its radial profiles are not significantly dissimilar to those of model~2.
To fit the model data, we use the Marquardt-Lovenberg algorithm. Because the fitting 
functions~(\ref{func1}) and (\ref{func2}) have
too many free parameters, this iterative algorithm takes to many iterations and often does not 
converge to reliable fits. Therefore, we vary manually the values of $n_1$, $n_2$, and $a$ 
until the best agreement with the model data is achieved and
obtain the best-fit values for $\Sigma_0$, $T_0$, $m$, and $b$. 
In both models, $T_{\rm bg}$ is set to 10~K.

We note that Equation~(\ref{func1}) is meant to approximate the gas surface density 
distribution of the disk only, while Equation~(\ref{func2}) does that for the midplane 
temperature of both the disk and the envelope. In addition, we fit the disk surface density 
only for $r\ge20$~AU and exclude the inner disk regions at $r<20$~AU, where a local 
peak/flattening in $\overline \Sigma$ is 
often seen. This feature is hardly present in the corresponding radial profiles of $\overline T$.
This local maximum or flattening in $\overline \Sigma$ 
is seen in many multidimensional numerical simulations of circumstellar disks with accretion
onto a forming star \citep[e.g.][]{Laughlin94,Krumholz07, Kratter10}. In our numerical
simulations this feature is likely caused by an absorbing inner computational boundary---the disk
material is allowed to freely flow through the sink cell but is not allowed to flow out of it. 
Indeed, we ran a few models
with a smaller sink cell ($r_{\rm sc}$=2~AU) and found that the position of the peak in 
$\overline\Sigma$ was always located a few~AU away from the inner computational boundary. 

\begin{table}
\begin{center}
\caption{Best fit parameters to radial profiles of $\overline \Sigma$ and $\overline T$}
\label{table2}
\begin{tabular}{ccccccccc}
\hline\hline
Model & Age & $\Sigma_0$ & $n_1$ & $a$ & $m$ & $T_0$ & $n_2$ & $b$   \\
\hline
1  & 0.1 & 56 & 0  & 40  & 1.8  & 138 & 0.45 & 1475  \\
1  & 0.3 & 35 & 0  & 22  & 0.8  & 110 & 0.45 & 900   \\
1  & 0.5 & 21 & 0  & 14  & 0.6  & 105 & 0.45 & 820   \\
1  & 0.7 & 14 & 0  & 13  & 0.5  & 103 & 0.45 & 790   \\
2  & 0.1 & 41150 & 1.7 & 150 & 1.0 & 290 & 0.5 & 3050  \\
2  & 0.3 & 10500 & 1.5 & 500 & 2.3 & 170 & 0.45 & 2290 \\
2  & 0.5 & 1870  & 1.2 & 550 & 2.5 & 145 & 0.45 & 1615 \\
2  & 0.7 & 575   & 1.0 & 550 & 2.5 & 130 & 0.45 & 1370 \\
2  & 0.7 & 31    & 0   & 80  & 0.8 & 130 & 0.45 & 1370 \\
 \hline
\end{tabular} 
\tablecomments{Disk age is in Myr, $\Sigma_0$ in g cm$^{-2}$, $T_0$ in Kelvin, and $a$ and $b$
are in AU.}
\end{center}
\end{table} 

The resulting best-fit parameters to model radial profiles of $\overline \Sigma$ and $\overline T$ 
in models~1 and 2 are listed in Table~\ref{table2}. We find that the radial profiles
of  $\overline \Sigma$ in model~1 are best fitted by a sole exponential 
function ($n_1=0$) rather than by the product of a power-law ($n_1\ne 0$)
and an exponential function. Indeed, the black solid lines in the top panels of Figure~\ref{fig3} 
show the lack of a distinct slope in the $\log~\overline\Sigma$ -- $\log~r$ space. Viscosity-dominated
polytropic disks are characterized by radial gas surface density profiles of similar form \citep{VB09b},
suggesting that the disk in model~1 is shaped by viscous torques rather than by 
gravitational ones. The similarity solution for a viscous disk with 
$\nu\equiv\alpha c_{\rm s} Z \propto r^{+1}$ predicts the inverse proportionality of 
$\overline \Sigma$ with radius \citep{Hartmann98}. The intermediate disk regions are indeed 
characterized by the ``classic'' $\overline \Sigma \propto r^{-1}$ scaling, which can be expected 
from the radial temperature dependence $\overline T\propto r^{-0.5}$, typical for flared disks, 
and from the radial vertical scale height dependence $Z\propto r^{1.25}$ obtained in 
detailed vertical disk structure models of \citet{DAlessio99}. At the same time,
the inner disk regions in model~1 show a shallower than $\overline\Sigma \propto r^{-1}$ scaling, while
the outer disk regions demonstrate a notably steeper scaling with radius. This steepening may 
partly be caused by flattening of the corresponding $\overline T$ profiles, 
making the sound speed $c_{\rm s}$ virtually independent of radius at large radii, 
and partly by disk flaring, which in our numerical simulations is described by a steeper 
than D'Alessio et al.'s dependence of $Z$ on radius at large radii \citep[see figure 7 in][]{Vor09b}.

The radial profiles of $\overline \Sigma$ in model~2 (and model~3) show a mixed behaviour. At the early
evolution $t\la 0.5$~Myr, they are best fitted by the product of a power-law and exponential function.
In the subsequent evolution, however, $\overline \Sigma$ begins to approach a pure exponential profile.
To demonstrate this trend, we fit the radial profile of $\overline \Sigma$ at $t=0.7$~Myr by both 
the product of a power-law and an exponential function ($n_1\ne0$, red dashed line) and also by a 
pure exponential function ($n_1=0$, blue dashed line). It is obvious that the latter case is more favourable.
It is known that protostellar disks dominated by gravity are characterized by gas surface density 
profiles that scale as $\overline\Sigma \propto r^{-1.5}$ \citep{VB09b,Rice10}, while the radial 
profiles of 
$\overline \Sigma$ in viscosity-dominated disks  cannot be described by a power-law function with 
a distinct exponent \citep{VB09b}. This suggests that the 0.5-Myr-old disk in model~2 makes a smooth
transition from the gravity-dominated to the viscosity dominated stage.


A visual inspection of Figure~\ref{fig3} reveals that  protostellar disks diminish, cool, and 
expand radially outward with time. This trend is also reflected in the time behaviour of the 
fitting parameters.
For instance, in both models $\Sigma_0$ and $T_0$ decrease with time. Concurrently, the exponents $m$
in model~1 and $n_1$ in model~2 decline notably with time, indicating a progressive shallowing of 
the corresponding gas surface density profiles. More specifically, the exponent $n_1$ takes a value
of 1.7 in the 0.1-Myr-old disk  of model~2 (early embedded phase) and drops to 1.0 
in the 0.7-Myr-old disk (T Tauri phase). 
In this context, it is interesting to note that T Tauri-type disks in the 1.0-Myr-old Ophiuchus 
star forming regions reveal $\overline\Sigma$ gradients with a mean value of $n_1\approx0.9$ \citep{Andrews09}.
The parameter $b$ decreases with time, indicating that the thermal
balance in the disk outer regions becomes dominated by the external environment rather than by internal
causes such as viscous (or shock) heating or stellar irradiation. We note that while the radial 
profiles of $\overline \Sigma$ shallow with time, the slope of the corresponding radial profiles 
of $\overline T$, expressed by the exponent $n_2$, remains virtually independent of time.


\begin{figure}
  \resizebox{\hsize}{!}{\includegraphics{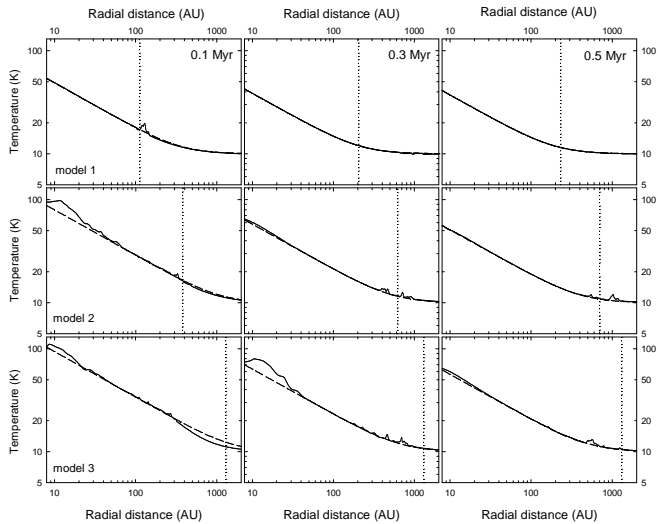}}
      \caption{Azimuthally-averaged gas midplane temperature $\overline T$ (solid lines) 
      and irradiation temperature $T_{\rm irr}$ (dashed lines) as a function of radius in model~1
      (top row), model~2 (middle row), and model~3 (bottom row). The left, middle, and right columns
      are the data obtained at $t=0.1$~Myr, $t=0.3$~Myr, and $t=0.5$~Myr after the disk formation, 
      respectively. The vertical dotted lines mark the position of the disk outer
      edge in each model.}
         \label{fig4}
\end{figure}

It is interesting to compare the azimuthally-averaged gas midplane temperature $\overline T$ 
with the irradiation temperature $T_{\rm irr}$ ($\phi$-independent) in our model 
protostellar disks at different stages of the evolution. The difference between the two 
values determines the extent to which the disk is locally non-isothermal in the vertical direction.
Figure~\ref{fig4} presents the radial profiles of $\overline T$ (solid lines) and $T_{\rm irr}$ 
(dashed lines) in model~1 (top row), model~2 (middle row), and model~3 (bottom row). 
The vertical columns correspond to different disk ages: $t=0.1$~Myr (left), $t=0.3$~Myr (middle), 
and t=0.5~Myr (right). The vertical dotted lines mark the position of the disk outer edge $r_{\rm d}$.

It is obvious that the midplane and irradiation temperatures
are virtually indistinguishable in the low-$M_{\rm cl}$ model~1, irrespective of the evolutionary stage,
indicating that low-mass disks are nearly isothermal in the  vertical direction. Only more massive disks
formed from cloud cores with $M_{\rm cl}\ga 0.9~M_\sun$ can show a notable negative 
vertical temperature gradient in the inner disk regions and this phenomenon is localized
to the early disk evolution. This temperature gradient is likely caused by viscous heating, 
which scales as $\Omega^2\propto r^{-3}$, thus operating preferentially in the disk inner regions 
\citep[e.g][]{Kratter08,Kratter10}, 
and also, to a lesser extent, by heating via shock waves with typical for our numerical simulations
Mach numbers of the order of 1--2. An example of the latter effect is seen in model~1 
near the outer edge of the 0.1-Myr-old disk (upper-left panel), where an accretion shock caused by 
the infalling envelope creates a local maximum in the gas midplane temperature. 
We note that in Figure~\ref{fig4} we have averaged the gas midplane temperatures in the azimuthal
direction, which may wash out large local azimuthal variations. We find that the gas midplane 
temperature in dense spiral arms and, especially, in massive fragments can be considerably 
greater than $T_{\rm irr}$ and reach several hundreds of Kelvin.

\begin{figure}
  \resizebox{\hsize}{!}{\includegraphics{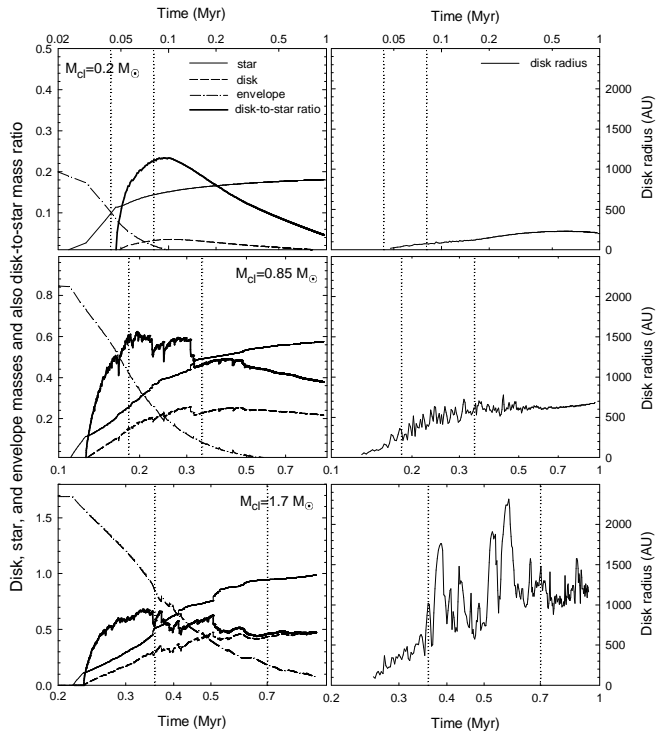}}
      \caption{{\bf Left column}. Time evolution of the star mass (solid lines), 
      disk mass (dashed lines), envelope mass (dash-dotted lines), and disk-to-star mass ratio (thick
      solid lines) in model~1 (top), model~2 (middle), and model~3 (bottom). {\bf Right column}. Time
      evolution of the disk radius $r_{\rm d}$ (solid lines) in model~1 (top), model~2 (middle), 
      and model~3 (bottom). vertical dotted lines mark the onset of the Class~I phase (left lines) and
      Class~II phase (right lines).}
         \label{fig5}
\end{figure}

\subsection{Time evolution of disk, stellar, and envelope masses}
Integrated disk and envelope masses can also give us an
insight into the rate at which protostellar disks evolve with time.
The left column in Figure~\ref{fig5} presents the stellar mass $M_\ast$ (solid lines), 
disk mass $M_{\rm d}$ (dashed lines), envelope mass $M_{\rm env}$ 
(dash-dotted lines), and also the disk-to-star mass ratio $\xi$ (thick solid lines)
as a function of time passed since the onset of cloud core collapse. 
The right column shows the disk outer radius $r_{\rm d}$.
The top row corresponds to model~1, while the middle and bottom rows
show the data for model~2 and model~3, respectively. The vertical dotted lines mark 
the onset of the Class~I (left) and Class~II (right) phases, which are determined using 
a classification breakdown of \citet{Andre93} based on the mass remaining in the envelope 
\citep[see][for details]{Vor10b}. Note that the masses and disk-to-star mass ratios 
in the left column are plotted at different scales.


The evolution of the disk-to-star mass ratio is most illustrative of changes occurring 
in protostellar disks with time. Soon after the disk formation in the Class~0 or early Class~I phase\footnote{In
fact, our disks form somewhat later than in the reality due to the use of the sink cell in our numerical
simulations. However, the difference in the formation times is only of order unity.},
$\xi$ quickly reaches a maximum value and then gradually declines with time. 
The peak disk-to-star mass ratio 
in model~1 is notably smaller ($ \xi\la 0.25$) than in model~2 ($\xi \la 0.6$) and model~3 
($\xi \la 0.7$). Higher $\xi$ favour gravitational instability and disk fragmentation, as 
reflected in disk images of Figure~\ref{fig1}.

It is evident that $\xi$ declines with time much faster in model~1 than in model~2 and 3, 
reflecting 
a progressively faster loss of disk material in lower-$M_{\rm cl}$ models due 
to accretion onto the star and, to a lesser extent, due to viscous expansion of the disk outer 
regions. This effect, however, is a consequence of the initial conditions. 
The surface density of our initial cores (see Equation~\ref{dens}) is close to the integrated 
column density of Bonnor-Ebert spheres with similar form ($r_{\rm out}/r_0=6$) and density 
enhancement ($A=1.2$), which implies that the low-mass cores
are characterized by smaller sizes and higher densities than their high-mass
counterparts (see Table~\ref{table1}).  
Such low-mass cores empty their mass reservoir faster, as seen from
the time behaviour of the envelope mass in Figure~\ref{fig5} and can be deduced from simple
free-fall-time arguments. This, in turn, means that 
disks in higher-$M_{\rm cl}$ models are replenished with the envelope material for a longer time,
leading to a slower decline in $\xi$. The situation could be reversed if we had formed 
more massive cores by taking a greater positive density enhancement $A$ but keeping the core size ($r_{\rm
out}$) and form fixed. In this case, more massive cores are expected to empty their mass 
reservoirs faster due to a higher infall rate onto the disk \citep[e.g.][]{Kratter10}.



Another interesting feature, related to the disk physics rather than to the initial core setup, 
is that the disk mass is steadily growing during the Class~0 phase  
and, to a lesser extent, in the Class~I phase and shows a trend
for decline only in the Class~II phase. This implies that the rate of mass transport 
of matter from the disk outer regions onto the star is smaller than the rate of mass loading onto 
the disk from the infalling envelope \citep[see also][]{Vor09a,Boley09a,Kratter10}. 
This phenomenon appears to be a necessary but not sufficient condition
for disk fragmentation and development of the burst mode of accretion \citep{VB10b}. 
Other factors such as insufficiently large disk radii and masses may suppress disk fragmentation
in low-$M_{\rm cl}$ models, as found in many previous studies 
\citep[e.g.][]{Matzner05,Rafikov05,Kratter10b}, even if
the rate of mass infall onto the disk is higher than the rate of mass accretion onto the star.

The time evolution of the stellar mass in the intermediate and upper-$M_{\rm cl}$ models
shows episodes of (almost) instantaneous increase, which are a manifestation of the accretion 
burst phenomenon
(note that the disk mass shows correlated drops). Such episodes are not present in the 
low-$M_{\rm cl}$ model.
The physical significance of such sharp increases in stellar mass for the early stellar evolution 
remains to be understood.

The time evolution of the disk radius $r_{\rm d}$ (right column in Figure~\ref{fig5}) 
reveals significant differences in models with distinct cloud core masses. In the 
$M_{\rm cl}=0.2~M_\sun$ model~1, the disk radius Steadily increases with time, reaches a maximum of
$r_{\rm d}$=240~AU at $t\approx0.5$~Myr and appears to gradually decline afterward, 
reflecting the ongoing stellar accretion and disk viscous dispersal. Most of the disk growth occurs
in the early Class~II phase due to viscous expansion.
On the other hand, the disk in model~2 ($M_{\rm cl}=0.85~M_\sun$) and, especially, in
model~3  ($M_{\rm cl}=1.7~M_\sun$) shows large scale radial pulsations, 
which are particularly strong in the Class~I and early Class~II phases. 

These episodes of disk expansion and contraction are byproducts of the burst
mode of accretion \citep{VB05,VB06,VB10b}. When a fragment forms in the disk via fragmentation 
of dense and
cold spiral arms, it is quickly driven onto the star via exchange of angular momentum with 
the arms. This fast migration produces a transient episode of disk expansion
due to conservation of angular momentum\footnote{Globally, angular momentum of the disk plus envelope
system is not conserved since we use the sink cell through which angular momentum is carried
away from the system by means of protostellar jets. However, mathematically, the total 
angular momentum (both that of the
system and that of the sink cell) is conserved, see tests in \citet{VB06}.}, 
which is followed by disk contraction. Such disk pulsations pursue as long as the disk 
is capable of fragmenting and are usually terminated soon after the Class~II phase ensues and the 
mass loading from the envelope diminishes. The $M_{\rm cl}=0.2$ model~1 is stable against 
fragmentation (see Figure~\ref{fig1}) and this explains why we do not see large scale disk 
pulsations in this model. 

Disk radial pulsations may be of great significance for the giant planet
formation mechanism via direct gravitational instability. Disk contraction leads to a 
transient episode of density increase, during which the disk may give birth to a  
set of protoplanetary embryos. Most of them would not survive through the EPSF and are either 
driven onto the star or dispersed. However, 
when mass loading from the natal cloud core diminishes and the main fragmentation phase ends,
the final disk contraction may give birth to a last and 
survivable set of gas giants on wide and relatively stable orbits \citep{VB10a}. 

\section{Disk structure and evolution along the line of increasing cloud core rotation rates}
\label{increnergy}
In this section, we study the structure and time evolution of protostellar disks
formed via the gravitational collapse of cloud cores characterized by distinct 
rotational energies. For this purpose, we 
consider three cloud cores with the ratio of the rotational to gravitational energy
$\beta=2.8\times 10^{-3}$ (model~4), $\beta=5.6\times 10^{-3}$ (model~2), and $\beta=2.3\times
10^{-2}$ (model~5). The ratio $\beta$ is varied by increasing the rotational energy $E_{\rm rot}$,
while the other model parameters, in particular the cloud core masses, are kept identical.

\begin{figure*}
 \centering
  \includegraphics[width=15cm]{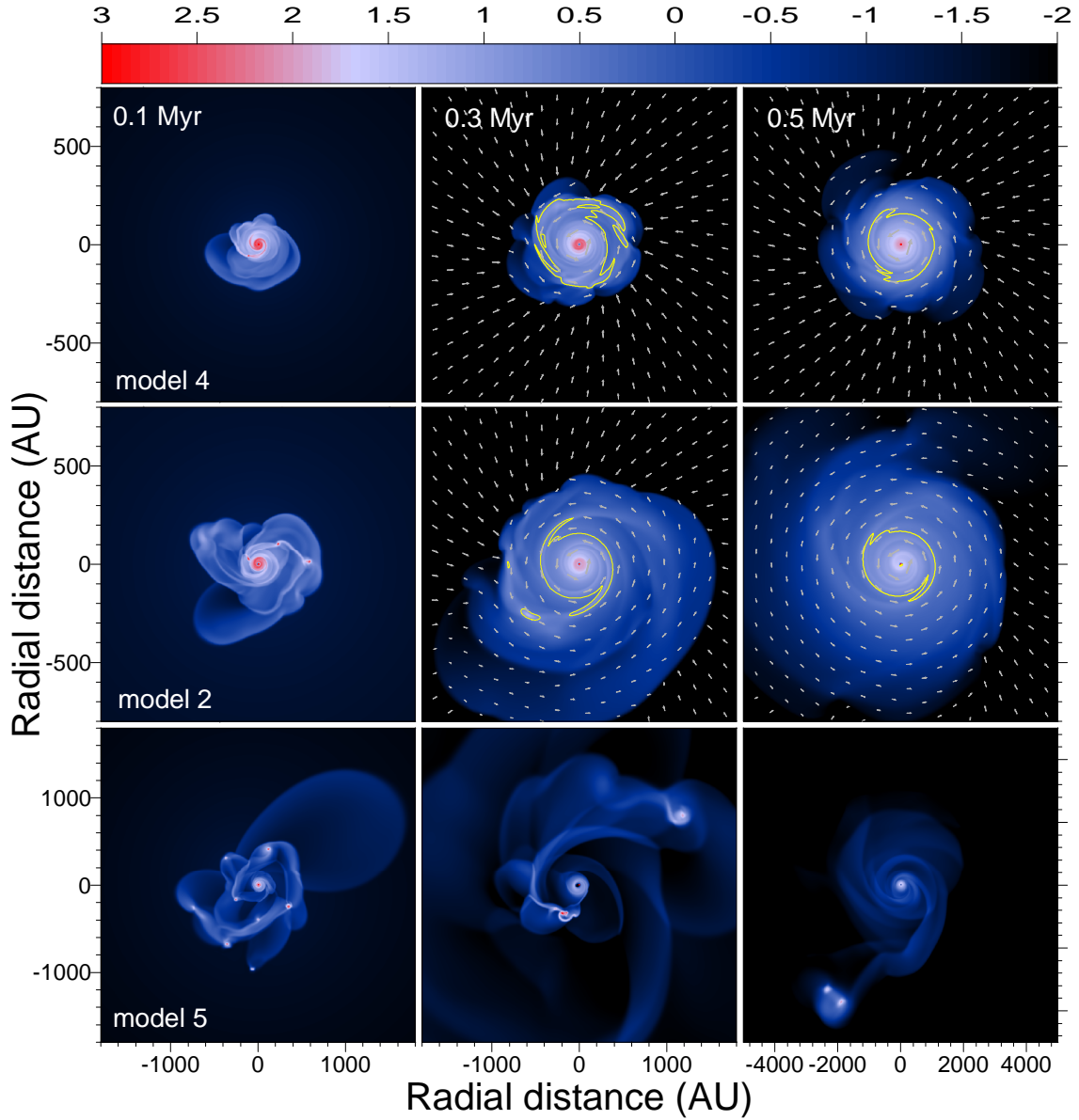}
      \caption{Gas surface density (in g~cm$^{-2}$, log units) in three models with distinct
      ratios of the rotational to gravitational energy $\beta$ and at distinct times $t$ after 
      the disk formation. In particular, the top row  shows disk images in the $\beta=2.8\times 10^{-3}$
      model~4, while the middle and bottom rows present those of the $\beta=5.6\times 10^{-3}$
      model~2 and $\beta=2.3 \times 10^{-2}$ model~5, respectively. Columns from left to right
      correspond to $t=0.1$~Myr, $t=0.3$~Myr, and $t=0.5$~Myr, respectively. 
      Superimposed on the disk
      images are the gas velocity fields (arrows) and  yellow contour lines delineating disk regions
      with the frequency-integrated optical depth $\tau\ge0.1$.}
         \label{fig6}
\end{figure*}

Figure~\ref{fig6} presents images of the gas surface density (in g cm$^{-2}$, log units) for model 4
(top row), model~2 (middle row), and model~5 (bottom row). The vertical columns from left to right correspond
to three distinct evolution times since the disk formation: $t=0.1$~Myr, $t=0.3$~Myr, and $t=0.5$~Myr.
The meaning of arrows and yellow contour lines is the same as in Figure~\ref{fig1}. We note the spatial
scale increases from  $800 \times 800$~AU for models~4 and 2 
to $1800 \times 1800$~AU  (at $t=0.1$~Myr and $t=0.3$~Myr) and $5000\times 5000$~AU 
(at $t=0.5$~Myr)  for model~5.

Figure~\ref{fig6} reveals similarities and also remarkable differences in the time evolution 
of protostellar disks formed from cloud cores of distinct $\beta$. While protostellar disks 
in all three models exhibit 
gravitational instability, which is strongest in the early evolution but notably 
weakens with time, only the $\beta=2.3\times10^{-2}$ model~5 shows vigorous disk fragmentation, 
with some of the fragments having survived to the late evolution stage.
The other two models with lower values of $\beta$ experienced disk fragmentation only 
in the very early stage ($t\la 0.1$~Myr, upper-left and upper-middle panels). However, the 
fragments did not survive, being either dispersed due to insufficient mass and/or numerical 
resolution or accreted onto the central star. 
By $t=0.5$~Myr, the high-$\beta$ model~5 appears to have formed a binary (or triple?) 
system, with masses of the primary and secondary being $0.45~M_\sun$  and $0.1~M_\sun$, respectively.
In fact, the latter value is an upper limit on both the secondary companion and its disk and the 
companion itself may end up as an upper-mass brown dwarf. The separation between the primary and secondary
was about 400~AU in the 0.3-Myr-old disk and it has increased by an order of magnitude in the
0.5-Myr-old disk, implying that the system may eventually break up and give birth to a rogue
upper-mass brown dwarf. 
Figures~\ref{fig1} and \ref{fig6} demonstrate that binary/multiple 
systems with {\it low} mass ratios can form
as a result of {\it disk} fragmentation and possible progenitors are cloud cores of 
sufficiently high original mass {\it and/or} sufficiently high rotation energy.
The latter requirement for disk fragmentation was also reported by \citet{Rice10}.
High-$\beta$ cores seem to form binaries/multiples at a notably larger spatial separation. 

Our present and previous numerical simulations \citep{VB06,VB10b} reveal that
most fragments migrate radially inward and (possibly) onto the star, while only a few 
migrate outward. 
Fragments migrate preferentially inward because they form within spiral arms and 
loose their angular momentum via gravitational
interaction with part of the arm located at a larger radial distance. In turn, this 
element of the arm experiences a 
positive torque and moves outward (see figure~5 in \citet{VB06} and animation at www.astro.uwo.ca/$\sim$vorobyov).
Even if a fragment forms at the very end of the arm, a continuing infall of the envelope material
exerts a negative torque on this fragment, driving it radially inward. This means that
only those fragments that form near the disk outer edge and in the final stages of the EPSF
(when infall from the envelope diminishes) may ultimately survive and either 
slowly migrate outward (bottom row in Figure~\ref{fig6}) or stabilize on orbits of order 100~AU
\citep{VB10a}.


The growing susceptibility of protostellar disks to fragmentation along the line of increasing 
$\beta$ can be  explained by the fact that the higher-$\beta$ cores form disks 
of larger size and greater disk-to-star mass ratio $\xi$. 
Both effects are mostly caused by the associated increase 
in the centrifugal radius $r_{\rm cf}=\Omega^2 r^4/G M(r)$ of a gas layer originally located 
at the radial distance $r$ from the center. 
In particular, the effect of disk radius on fragmentation has been addressed by many
authors \citep[e.g.][]{Rafikov05, Matzner05, Kratter10b, VB10b} who have found that
fragmentation is suppressed in small-size disks due to either 
stellar irradiation or viscous heating.


\begin{figure}
  \resizebox{\hsize}{!}{\includegraphics{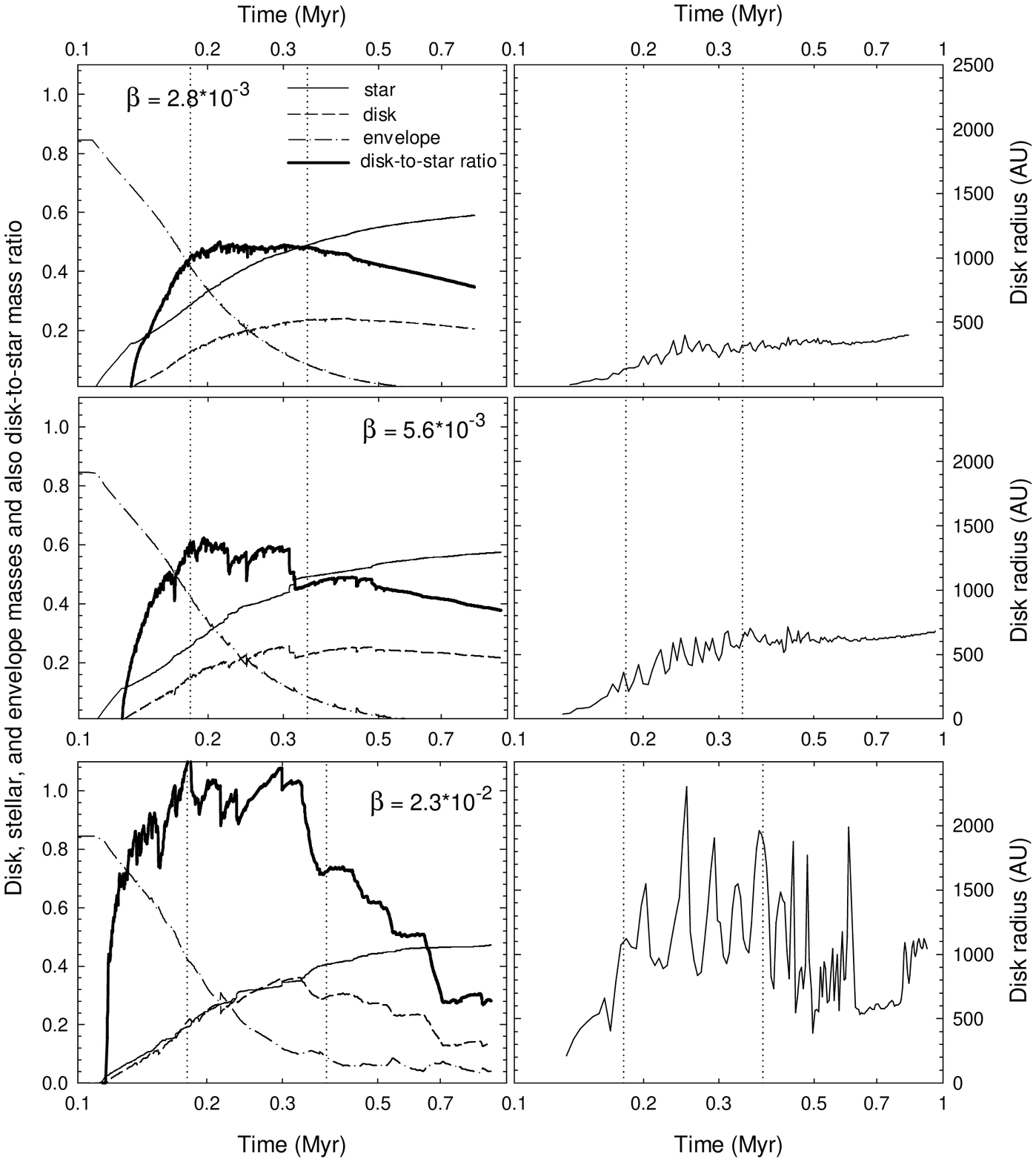}}
      \caption{{\bf Left column}. Time evolution of the star mass (solid lines), 
      disk mass (dashed lines), envelope mass (dash-dotted lines), and disk-to-star mass ratio (thick
      solid lines) in model~4 (top), model~2 (middle), and model~5 (bottom). {\bf Right column}. Time
      evolution of the disk radius $r_{\rm d}$ (solid lines) in model~4 (top), model~2 (middle), 
      and model~5 (bottom). Vertical dotted lines mark the onset of the Class~I phase (left lines) and
      Class~II phase (right lines).}
         \label{fig7}
\end{figure}

The left column in Figure~\ref{fig7} presents the time evolution 
of disk masses (dashed lines), stellar masses (solid lines), envelope masses (dash-dotted lines), 
and disk-to-star mass ratios (thick solid lines), while the right
column shows the time evolution of disk radii (solid lines). 
In particular, the top row presents the data for the $\beta=2.8\times 10^{-3}$ model~4, 
while the middle and bottom rows correspond to the $\beta=5.6\times 10^{-3}$ model~2 
and $\beta=2.3\times 10^{-2}$ model~5, respectively. The vertical dotted lines mark 
the onset of the Class~I (left) and Class~II phases (right) in each model. 

In all three models, the disk-to-star mass ratio $\xi$ quickly grows with time in the 
Class~0 phase, attains a maximum value in the late Class~0 or early Class~I phase, and declines 
steadily afterwards. In the low- and intermediate-$\beta$ models, the disk mass is always lower 
than that of the central star, with maximum $\xi=0.5$ (model~4) and $\xi=0.6$ (model~2).  
On the other hand, the disk mass in the high-$\beta$ model~5  exceeds episodically that of the 
star in the Class~I phase but drops quickly to $\xi\la 0.4$ in the subsequent evolution when 
the binary system starts to emerge. This suggests that protostellar disks with mass comparable 
to or greater than that of the host star must be statistically rare and this phenomenon quickly 
resolves into a binary/multiple system. 
A similar effect was reported in numerical hydrodynamics simulations of \citet{Kratter10}.
Moreover, such massive disks must be difficult to observe due to the obscuration of light 
by natal envelopes that are still of substantial mass in the Class~I phase. 
Protostellar disks formed from higher-$\beta$ 
cores undergo radial pulsations of a notably greater amplitude.
This increase is caused by a growing strength of gravitational instability and fragmentation. 
In all three models, the radial pulsations appear to subside with time.

\begin{figure*}
 \centering
  \includegraphics[width=15cm]{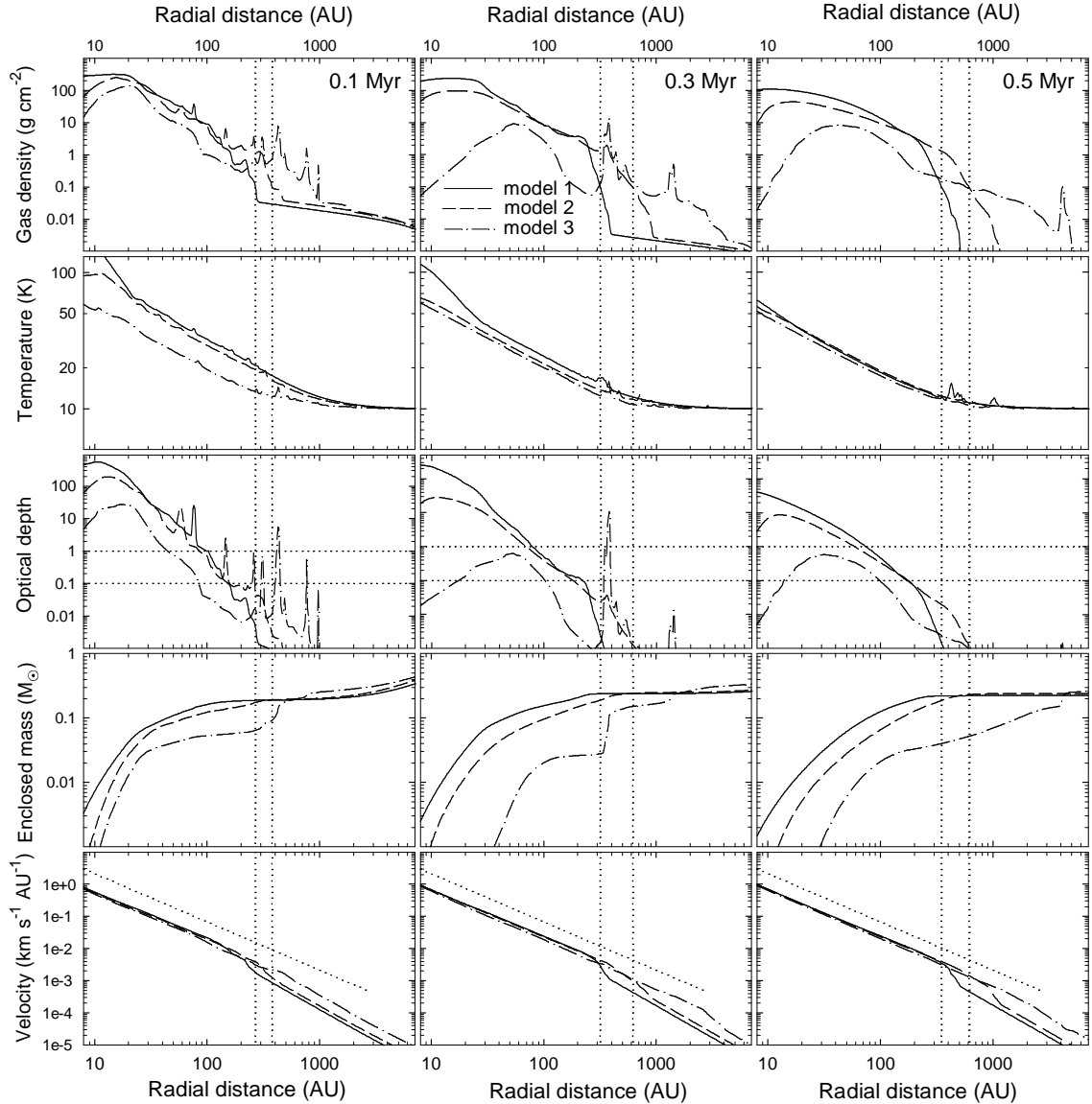}
      \caption{Azimuthally-averaged radial profiles of the gas surface density
      $\overline \Sigma$ (first row), gas midplane temperature $\overline T$ (second row), 
      frequency-integrated optical depth $\overline \tau$ (third row), and angular velocity 
      $\overline \Omega$ (fifth row). The forth row presents the enclosed mass $M(r)$ as a function
      of radius. The vertical columns from left to right show the data at three different times after the
      disk formation: $t$=0.1~Myr (left), $t$=0.3~Myr (middle), and $t=0.5$~Myr (right).
      The solid, dashed, and dash-dotted lines present the data for the $\beta=2.8\times 10^{-3}$ model~4,
      $\beta=5.6\times 10^{-3}$ model~2, and $\beta=2.3\times 10^{-2}$ model~5, respectively. 
      The vertical dotted lines mark the position
      of the disk outer edge in model~4 (left lines) and model~2 (right lines). The meaning of other
      dotted lines is the same as in Figure~\ref{fig1}.}      
         \label{fig8}
\end{figure*}

Finally, in Figure~\ref{fig8} we consider the azimuthally-averaged radial profiles 
(from top to bottom row) of the gas surface density, midplane gas temperature, optical depth, integrated
mass, and angular velocity at $t=0.1$~Myr (left column), $t=0.3$~Myr (middle column), and $t=0.5$~Myr
(right column) after the disk formation. In particular, the solid, dashed and dash-dotted lines correspond
to model~4, model~2, and model~5, respectively. The meaning of the dotted lines is the same as in Figure~\ref{fig3}.

A visual inspection of Figure~\ref{fig8} reveals that protostellar disks formed from cloud 
cores of increasingly higher $\beta$ are characterized by {\it lower} azimuthally-averaged 
gas surface density $\overline \Sigma$, midplane temperature $\overline T$, and optical depth 
$\overline \tau$. At the same time, both the disk size and mass increase along the line 
of increasing $\beta$. Lower $\overline \Sigma$ do not prevent higher-$\beta$ disks from fragmenting,
as Figure~\ref{fig6} demonstrates. On the contrary, these disks are more susceptible to fragmentation,
owing to their larger size and lower gas temperature. The latter effect is caused by lower stellar 
irradiation fluxes at larger radii and lower viscous heating \citep[e.g.][]{Matzner05,Kratter10b}. 
A substantial fraction of protostellar disks in 
the low- and intermediate-$\beta$ models~4 and 2 are optically thick, suggesting that the mass 
of such disks may be systematically underestimated using conventional observational techniques.
Sharp jumps in the integrated mass of model~5 at 400~AU (0.1-Myr-old disk), 1200~AU (0.3-Myr-old disk),
and 4200~AU (0.5-Myr-old disk) reflect the formation and outward drift of a sub-stellar companion. 



\section{Conclusions}
\label{conclude}
Using numerical hydrodynamics simulations in the thin-disk approximation, we model 
the formation and evolution of protostellar 
disks in the embedded phase of star formation when the disks are exposed to intense mass loading 
from natal cloud cores and are difficult to probe using conventional observational techniques.
The key physical processes that are taken into consideration in our modeling include: disk self-gravity,
viscous and shock heating, stellar and background irradiation, radiative cooling from the disk surface,
and also self-consistent accretion of gas from the disk onto the forming star and from the infalling
envelope onto the disk. We scrutinize the structure and time evolution of protostellar disks 
formed from five model cloud cores with distinct initial masses in the 0.2--1.7~$M_\sun$ 
range and ratios of the rotational to gravitational energy $\beta$ in the 
0.28--2.3$\times 10^{-2}$ range.  We find the following.

\begin{itemize}
\item The time evolution of embedded protostellar disks proceeds from a conspicuously non-axisymmetric
state toward a more regular, axisymmteric appearance. Simultaneously, the gas surface 
density reduces and disks expand radially outward with time. 
However, the time scale for this transformation is faster in lower-mass disks formed 
from cloud cores of smaller mass. For instance, disks formed from cloud cores
of $M_{\rm cl}=0.2~M_\sun$ are virtually axisymmetric at $t=0.3$~Myr after their formation, 
while more massive disks formed from cloud cores of $M_{\rm}=0.85~M_\sun$ retain a notable 
diffuse spiral structure at $t=0.5$~Myr. In the non-axisymmetric state, massive 
disks are often lopsided and the gas flow exhibits non-circular motions, 
indicating local deviations from a circular rotation caused by spiral arms and forming fragments. 

\item Disk fragmentation is seen in most models. However, most of the fragments do not survive through
the embedded phase and are either destroyed or driven onto the forming star. Only models with sufficiently
high $M_{\rm cl}$ or $\beta$ (for instance, $M_{\rm cl}=1.7~M_\sun$ and $\beta=5.6\times 10^{-3}$ {\it
or} $M_{\rm cl}=0.85~M_\sun$ and $\beta=2.3\times 10^{-2}$)
reveal the formation of wide-separation binary/multiple systems with low mass ratios and 
(possibly) brown dwarf companions. In particular, systems formed from cloud cores with 
$\beta\ga 2.3\times 10^{-2}$ may ultimately break up, delivering freely wandering 
brown dwarfs to the natal star forming region.
These rogue brown dwarfs may even possess some minidisks, the properties of which remain 
to be studied. 

\item The fact that only sufficiently massive cores with sufficiently high initial rotational 
energy $E_{\rm rot}$ form binary systems
can in part account for a lower binary fraction found in stars of lower mass. Low-mass stars 
tend to form from low-mass cores and only a small fraction of these cores with high enough 
$E_{\rm rot}$ can give birth to a binary system. As the core mass grows (and so does the resulting 
mass of the star), more and more cores can form binary systems because the requirement of 
sufficiently high initial rotational energy is relaxed along the line of increasing core mass.


\item Embedded protostellar disks of {\it equal} age formed
from cloud cores of greater mass (but equal $\beta$) 
are generally characterized by higher gas surface densities, 
higher midplane temperatures, larger radii, and greater masses. 
On the other hand, protostellar disks formed from cloud cores of higher $\beta$ 
(but equal $M_{\rm cl}$) are generally thinner and colder but larger and more massive. 

\item
The trend of more massive cloud cores to form denser and hotter disks diminishes for 
$M_{\rm cl}\ga0.9~M_\sun$ owing to the onset of vigourous gravitational instability and 
fragmentation.  Gravitational torques efficiently transport matter in the form of fragments from 
the disk outer regions (where most of the envelope material lands and fragmentation takes place) 
onto the star, preventing the disk from growing in density and
bringing it back to the border of stability. However, the disk continues to 
grow in size (and mass) due to infall of the envelope material with high angular momentum
(which may lead to another fragmentation episode). This process continues until the 
envelope is depleted of matter.


\item The positive difference between the midplane and irradiation temperatures 
is significant only in the inner disk regions ($r\la 50$~AU),  
near the disk's outer edge and dense fragments, where it can reach a factor of two or more. 
The degree of 
vertical non-isothermality becomes progressively smaller in disks formed from cloud cores of smaller
mass and as the disks evolve.

\item Protostellar disks in the embedded phase (especially in the Class~I phase) 
show radial pulsations, the amplitude of which increases along the line of increasing $M_{\rm cl}$
and $\beta$. These pulsations are caused by gravitational instability and, to a greater extent,
by disk fragmentation and accretion of fragments onto the protostar. The amplitude of radial pulsations
appears to diminish in the Class~II phase when the envelope clears.

\item The disk-to-star mass ratio $\xi$ increases along the line of increasing $M_{\rm cl}$
and/or $\beta$. A maximum value of $\beta$ of order unity is achieved in single stars formed from 
cloud cores with high rates of rotation $\beta\ga 2.3\times 10^{-2}$. However, such massive
disks are short-lived and by the end of the embedded phase they quickly evolve into a binary/multiple
system.

\item A substantial fraction of a protostellar disk in the embedded phase may be optically thick,
implying that observationally inferred disk masses may be underestimated.

\end{itemize}

We also provide useful approximation formula to the radial profiles of the gas surface density
and midplane temperature at different stages of the evolution and for disks of different mass. 
We did not see the formation of planetary-mass objects in protostellar disks 
as in numerical simulations with a barotropic equation of state by \citet{VB10a}. However, we
hope to see them forming in our future simulations because some of our low-mass fragments seem 
to disperse due to the lack of numerical resolution at large radii where our logarithmic grid diverges.
Finally, it is worth noting that the results may be somewhat dependent 
on the chosen angular momentum and density profiles. We plan to vary the initial profiles 
in a future study.



\acknowledgements
The author is thankful to the anonymous referee for useful comments and suggestions that helped to
improve the paper. The author gratefully acknowledges present support 
from an ACEnet Fellowship. Numerical simulations were done 
on the Atlantic Computational Excellence Network (ACEnet),
on the Shared Hierarchical  Academic Research Computing Network (SHARCNET),
and at the Center of Collective Supercomputer
Resources, Taganrog Technological Institute at Southern Federal University.
This project was also supported by RFBR grant 10-02-00278 and by Ministry of Education grant 
RNP 2.1.1/1937.

\end{document}